 \documentclass[11pt,twoside]{article}
\newcommand{\barefootnote}[1] {%
  \begingroup
    \renewcommand{\thefootnote}{}
    \footnotetext{#1}
    \renewcommand{\thefootnote}{\arabic{footnote}}
  \endgroup
}

\renewcommand{\title}[1] {%
  \begingroup
    \begin{center}
      \vspace{0.4in}
      \bf\huge
      \addtolength{\baselineskip}{5mm}
      #1
    \end{center}
  \endgroup
}

\newcommand{\url}[1] {%
  \barefootnote{%
    {\small e-print archive: }
    {\texttt http://xxx.lanl.gov/#1}
  }
}

\renewcommand{\author}[1] {%
  \begingroup
    \begin{center}
      \vspace{0.4in}
      \bf
      #1
      \vspace{0.2in}
    \end{center}
  \endgroup
}

\newcommand{\address}[1] {%
  \begingroup
    \begin{center}
      #1
    \end{center}
  \endgroup
}

\newcommand{\addressemail}[1] {%
  \begingroup
    \begin{center}
      \vskip-\baselineskip
      #1
    \end{center}
  \endgroup
}

\newcounter{secondpage}
\newcounter{gfirstpage}
\newcounter{glastpage}
\catcode`\@=11
\def\newsymbol#1#2#3#4#5{\let\next@\relax%
 \ifnum#2=\@ne\else%
 \ifnum#2=\tw@\let\next@\msyfam@\fi\fi%
 \mathchardef#1="#3\next@#4#5}
\def\mathhexbox@#1#2#3{\relax%
 \ifmmode\mathpalette{}{\m@th\mathchar"#1#2#3}r
 \else\leavevmode\hbox{$\m@th\mathchar"#1#2#3$}\fi}
\def\hexnumber@#1{\ifcase#1 0\or 1\or 2\or 3\or 4\or 5\or 6\or 7\or 8%
\or 9\or A\or B\or C\or D\or E\or F\fi}
\font\tenmsy=msbm10
\font\sevenmsy=msbm7
\font\fivemsy=msbm5
\newfam\msyfam
\textfont\msyfam=\tenmsy
\scriptfont\msyfam=\sevenmsy
\scriptscriptfont\msyfam=\fivemsy
\edef\msyfam@{\hexnumber@\msyfam}
\def\Bbb#1{\fam\msyfam\relax#1}
\catcode`\@=\active
\newtheorem{theorem}{Theorem}[section]
\newtheorem{proposition}[theorem]{Proposition}
\newtheorem{lemma}[theorem]{Lemma}

\newcommand{\proof}{{\noindent \it Proof:\ }}
\newcommand{\qed}{\hfill $QED$\par\medskip}
\newcommand{\BR}{{{\Bbb R}^3}}
\newcommand{\ct}{{\Bbb C}^2}
\newcommand{\BC}{{\Bbb C}}
\newcommand{\RR}{{{\Bbb R}}}

\newcommand{\xo}{x\otimes\Omega}

\newcommand{\LR}{{L^2({\Bbb R}^3)}}

\newcommand{\nm}{(\kj)}

\newcommand{\fff}{{\ff}}
\newcommand{\ff}{{\cal F}}

\newcommand{\fffr}{{\cal F}_{\rm real}}

\newcommand{\hann}{{1\! /\!2}}
\newcommand{\hpz}{H_{p0}}
\newcommand{\res}{\left(1+\frac{t}{n}\hpz\right)^{\!\!-\hann}}
\newcommand{\ress}{\left(1+\frac{t}{n}\hpz\right)\f}
\newcommand{\zes}{\left(z-\hpz\right)^{\!\!-\hann}}
\newcommand{\zess}{\left(z-\hpz\right)\f}

\newcommand{\f}{^{-1}}

\newcommand{\kj}{k_1,j_1,...,k_n,j_n}
\newcommand{\hpk}{H_{p-k}}

\newcommand{\ffff}{{\cal F}_{\rm fin}}

\newcommand{\djj}{\delta_{jj'}}

\newcommand{\jj}{\sum_{j=1,2}}

\newcommand{\sumi}{\sum_{i=1}^n}

\renewcommand{\aa}{e}
\newcommand{\az}{\aa^\ast} 
\renewcommand{\sp}{\Sigma_{\rm el}}
\newcommand{\epp}{E_{\rm el}}
\newcommand{\pel}{P_{\rm el}} 

\newcommand{\gap}{\sp-\epp}

\newcommand{\mm}{|\aa|}
\newcommand{\at}{\aa^2}
\newcommand{\add}{a^{\ast}} 
\newcommand{\aaa}{a}

\newcommand{\Av}{A_{\varphi}}

\newcommand{\Bv}{B_{\varphi}}

\newcommand{\mmm}{\sum_{\mu=1,2,3}}

\newcommand{\ass}{a^{\sharp}}
\newcommand{\half}{\frac{1}{2}}

\newcommand{\kx}{k\cdot x}

\newcommand{\hhh}{{\cal H}}

\newcommand{\lk}{\left(}

\newcommand{\rk}{\right)}
\newcommand{\lkk}{\left\{}
\newcommand{\rkk}{\right\}}

\newcommand{\hf}{{H_{\rm f}}
}\newcommand{\hv}{H_V} 
\newcommand{\dm}{\omega_0}
\newcommand{\pf}{{P_{\rm f}}}
\newcommand{\nf}{{N_{\rm f}}}

\newcommand{\hp}{H_p}

\newcommand{\infk}{\inf_{k\in\BR}}

\newcommand{\hi}{H_{p{\rm I}}}

\newcommand{\hib}{(\hi-E(p))}
\newcommand{\hiw}{\widetilde{{H_{\rm I}}}}

\newcommand{\hii}{H_{p{\rm II}}} 
\newcommand{\hiiw}{{\widetilde{H}_{p{\rm II}}}} 
 
\newcommand{\rpof}{{\cal O}_{\rm real}} 
\newcommand{\epr}{E(\p)}

\newcommand{\tr}{{\rm Tr}}

\renewcommand{\t}{\theta}
\newcommand{\ee}{\eta(e)} 
\newcommand{\co}{c_0} 
\newcommand{\cco}{c_0(\aa)}

\newcommand{\D}{\Delta}
\newcommand{\pa}{H_{\rm el}}

\newcommand{\s}{{\rm Spec}} 
\newcommand{\si}{\sigma} 

\newcommand{\BB}{\widetilde{B}_{\varphi}}

\newcommand{\vp}{\widehat{\varphi}}
\newcommand{\vP}{\hat{p}}
 
\newcommand{\hz}{{{\Bbb Z}_{1\!/\!2}}}
\newcommand{\pj}{\vP\!\cdot\! J} 

\newcommand{\hk}{\widehat{k}}

\newcommand{\p}{p}

\newcommand{\jf}{J_{\rm f}}
\newcommand{\ssf}{S_{\rm f}}

\newcommand{\ejt}{e^{i \t \n\cdot (\jf+\ssf)}}
\newcommand{\emjt}{e^{-i \t \n\cdot (\jf+\ssf)}}
\newcommand{\eejt}{e^{i \t \n\cdot J}}
\newcommand{\eemjt}{e^{-i \t \n\cdot J}}

\newcommand{\n}{\vec{n}}

\newcommand{\pc}{\p_{\rm c}}

\newcommand{\ppo}{\Omega_+}
\newcommand{\ppoo}{\Omega_\pm}

\newcommand{\mmo}{\Omega_-}

\newcommand{\eq}[1]{\begin{equation}
\label{#1}}
\newcommand{\en}{\end{equation}}
\newcommand{\bl}[1]{\begin{lemma}
\label{#1}}
\newcommand{\el}{\end{lemma}}

\newcommand{\bt}[1]{\begin{theorem}
\label{#1}}
\newcommand{\et}{\end{theorem}}

\newcommand{\bi}{\begin{description}}
\newcommand{\ei}{\end{description}}

\newcommand{\bp}[1]{\begin{proposition}
\label{#1}}
\newcommand{\ep}{\end{proposition}}
\newcommand{\PP}{(p-\pf-\aa \Av)}

\newcommand{\pg}[1]{\sum_{\phi\in \{\phi_i\}}(\phi,{#1}\phi)}

\renewcommand{\P}{\pg} 
\renewcommand{\pg}{P_{\rm g}} 

\newcommand{\sa}{\frac{\vp(k)}{\sqrt{2\omega(k)}}}

\newcommand{\kak}[1]{(\ref{#1})}

\newcommand{\ehtt}{e^{-t(\hp-\epr)}}
\newcommand{\gr}{\psi_{\rm g}} 
\newcommand{\gri}{\psi_{\rm g+}} 
\newcommand{\grii}{\psi_{\rm g-}} 
\newcommand{\griii}{\psi_{\rm g\pm}} 

\newcommand{\fri}{\psi_{+}} 
\newcommand{\frii}{\psi_{-}} 
\newcommand{\friii}{\psi_{\pm}}

\newcommand{\qqq}{\int \lk {\omega(k)^{-2}}+1\rk |\vp(k)|^2 dk}
\newcommand{\qq}{\int \lk {\omega(k)^{-2}}+\omega(k)\rk |\vp(k)|^2 dk}

\renewcommand{\ln}{\lim_{n\rightarrow\infty}}
\newcommand{\lnk}{\ln\!\lim_{m\rightarrow\infty}}

\newcommand{\lime}{\lim_{\e\rightarrow 0}}

\newcommand{\e}{\epsilon}
 
\begin{document}
 \title{Ground state degeneracy of the Pauli-Fierz Hamiltonian with  
spin}
% 
% \url{hep-th/98111131}		%% Everything after http://xxx.lanl.gov/
% 
 \author{F. Hiroshima and H. Spohn} 
\makeatletter
 \address{Department of Mathematics and Physics\\ 
Setsunan University\\
Ikeda-naka-machi 17-8, Neyagawa\\
Osaka, Japan}
\addressemail{hiroshima@mpg.sestunan.ac.jp}
\address{\
 Technische Universit\"at M\"unchen \\
Zentrum  Mathematik \\
Arcisstr. 21\\ D-80290
 M\"unchen, Germany}
\addressemail{spohn@ma.tum.de}
\makeatletter
 \markboth{\it Degenerate ground states\ldots}{\it F. Hiroshima and H. Spohn} 
 \begin{abstract}
We consider an electron, spin $\hann$, minimally coupled to the 
quantized radiation field in the nonrelativistic approximation, a 
situation defined by the Pauli-Fierz Hamiltonian $H$. 
There is no external potential and $H$ fibers as $\int^\oplus \hp 
dp$ according to 
the total momentum $p$. 
We prove that the ground state subspace of $\hp$ is two-fold 
degenerate provided the charge $e$ and the total momentum $p$ are 
sufficiently small. 
We also establish that the total angular momentum of the ground state 
subspace is $\pm\hann$ and study the case of a confining external 
potential. 
 \end{abstract}

\setlength{\topmargin}{0in}
\setlength{\headheight}{9pt}%{0in}
\setlength{\textwidth}{5in}
\setlength{\textheight}{8in}
\setlength{\paperheight}{11in}
\setlength{\paperwidth}{8.5in}
\setlength{\parskip}{2ex}

\section{Introduction and main results} 
Nonrelativistic quantum electrodynamics predicts the anomalous 
magnetic moment ($g$-factor) of the electron with 
an error of less than $1\%$. 
One definition of the $g$-factor comes from the motion of the 
electron in a weak uniform external 
magnetic field. 
$g/2$ is then the ratio of the spin precession relative to the 
orbital precession. 
To provide a theoretical analysis of  such an experiment one has to 
show that 
the ground state band of the Hamiltonian is adiabatically decoupled 
from the spectrum of excitations and has to determine the effective 
Hamiltonian governing 
the motion in the ground state band. The details are given elsewhere 
\cite{gts}. 
They result in a nonperturbative microscopic definition of the 
$g$-factor, which, when computed to order $e^2$, yields $g/2=1.0031$ 
as compared to the accurate relativistic value of $g/2=1.0012$. 
One crucial input of  the analysis is the microscopic Hamiltonian at 
fixed total 
momentum to have 
an exactly two-fold degenerate ground state. 
In our paper we will, under suitable conditions, establish such a 
spectral property. 

Let us first introduce the Pauli-Fierz Hamiltonian for a single 
electron 
coupled to the quantized radiation field. 
By translation invariance the total momentum is conserved and 
the operator of interest will be the Pauli-Fierz Hamiltonian fibered 
with respect to the total momentum. 
The Hilbert space for the coupled system is 
$$\hhh=\LR\otimes \BC^2\otimes\fff.$$
$\LR$ is the Hilbert space for the translational degrees of freedom of 
the electron, 
multiplication by $x$ stands for the position, 
$-i\nabla_x$ for the momentum of the electron. 
$\BC^2$ is the Hilbert space for the spin,  $\si$, 
of the electron, 
where $\si=(\si_1,\si_2,\si_3)$ are the  spin $1/2$ Pauli spin 
matrices, 
$$
\si_1=\lk 
\begin{array}{cc}
0&1\\  
1&0
\end{array}\rk, \ \ \ 
\si_2=\lk 
\begin{array}{cc}
0&-i\\ 
i&0
\end{array}\rk,\ \ \ 
\si_3=\lk 
\begin{array}{cc}
1&0\\
0&-1\end{array}\rk. 
$$
$\fff$ is the symmetric Fock space for the photons given by 
$$\fff=\oplus_{n=0}^\infty \lk \LR\otimes \BC^2\rk_{\rm sym}^n,$$
where  $(\cdots)_{\rm sym}^n$ denotes the $n$-fold symmetric tensor 
product of $(\cdots)$. 
$\LR\otimes\BC^2$ will be identified with $L^2(\BR\times\{1,2\})$. 
The Fock vacuum is denoted by $\Omega$ and a vector $\psi\in\fff$ by 
$\psi=(\psi^{(0)},\psi^{(1)},...)$, $\psi^{(0)}=c\Omega$, $c\in\BC$. 
The photons live in $\BR$ and have  helicity $\pm1$. 
The photon field is thus represented by 
the two-component Bose field $a(k,j)$ on  $\fff$ with commutation 
relations 
$$[a(k,j),\add(k',j')]=\djj\delta(k-k'),$$
$$[a(k,j),a(k',j')]=0,\ \ \ [\add(k,j),\add(k',j')]=0,$$
$k,k'\in\BR$, $j,j'=1,2$. 
 The energy of the photons is given by 
\eq{01}
\hf=\jj  \int\omega(k)\add(k,j)\aaa(k,j)dk,
\en 
i.e. $\hf$ restricted to $(\LR\otimes \BC^2)_{\rm sym}^n$ is  
multiplication by 
$\sum_{j=1}^n\omega(k_j)$. 
Throughout units are such that $\hbar=1$, $c=1$. 
Physically 
$\omega(k)=|k|$. 
This  case is somewhat singular. To regularize,  the photons are 
assumed to have 
a small mass as 
\eq{10}
\omega(k)=\sqrt{|k|^2+m_{\rm ph}^2},\ \ \ m_{\rm ph}>0.
\en 
In fact we can allow for a more general class of dispersion 
relations. 
We assume $\omega:\BR\rightarrow \RR$ to be continuous with the 
properties 
\bi
\item[(A.1)]
$\infk \omega(k)\geq \dm>0,$
\item[(A.2)]
$\omega(k_1)+\omega(k_2)\geq \omega(k_1+k_2),$
\item[(A.3)] $\omega(k)=\omega(Rk)$ for an arbitrary  rotation $R$. 
\ei 

The quantized transverse vector potential is defined through 
$$\Av(x)= \jj \int \frac{\vp(k)}{\sqrt{2\omega(k)}} e_j(k)  \lk 
e^{-i\kx}
\add(k,j)+e^{i\kx}  a(k,j)\rk dk.$$
Here 
$e_1$ and $e_2$ are  polarization vectors which together with 
$\hk=k/|k|$ 
form a standard basis in $\BR$. 
$\varphi:\BR\rightarrow \RR$ is the  form factor which ensures an 
ultraviolet cutoff. 
It is assumed  to be rotational invariant, 
$\varphi(Rx)=\varphi(x)$ for an arbitrary rotation $R$, continuous, 
bounded with some decay at infinity, and normalized as 
$\int \varphi(x) dx=1$. We will mostly 
work with the Fourier transform $\vp(k)=(2\pi)^{-3/2}\int\varphi(x) 
e^{-i\kx} dx$. 
It satisfies 
(1) $\vp(Rk)=\vp(k)$ for an arbitrary  rotation $R$,
(2)  $\overline{\vp}=\vp$ for notational simplicity, 
(3) the normalization $\vp(0)=(2\pi)^{-3/2}$,  and 
(4) the decay 
$$\int\lk \omega(k)^{-2}+\omega(k)\f+1+\omega(k)\rk |\vp(k)|^2 
dk<\infty.$$
With these preparations the Pauli-Fierz Hamiltonian, including spin, 
is defined by 
\eq{11}
H=\frac{1}{2m}\lkk\si\cdot\lk-i\nabla_x-e\Av(x)\rk\rkk^2+\hf,
\en 
where $m$ is the bare mass and $e$ the charge of the electron.

Translations for the electron are generated by 
$-i\nabla_x$ and translations  for  the photon field by the field 
momentum 
\eq{02}
\pf=\jj \int k\add(k,j)\aaa(k,j)dk.
\en
By translation invariance of $H$ the total momentum 
$$p=-i\nabla_x+\pf$$
is thus conserved, 
$$[H,p]=0,$$
as  can be checked also directly. 
We unitarily transform $H$ such that the fibering with respect to $p$ 
becomes 
transparent. 
In momentum representation, momentum multiplication by $k$, 
position $i\nabla_k$, an element $\psi\in\hhh$ is written as 
$\psi^{(n)}(k,k_1,j_1,...,k_n,j_n)$ with values in $\BC^2$. 
For each $n$ let 
\eq{12}
U \psi^{(n)}(k,k_1,j_1,...,k_n,j_n)=
\psi^{(n)}(k-\sum_{j=1}^n k_j, k_1,j_1,...,k_n,j_n)
\en 
with inverse
\eq{13}
U\f  \psi^{(n)}(k,k_1,j_1,...,k_n,j_n)=
\psi^{(n)}(k+\sum_{j=1}^n k_j, k_1,j_1,...,k_n,j_n).
\en 
Clearly $U$ is unitary. We set 
$$\Av=\Av(0)$$
as an operator on $\fff$. Similarly for the quantized magnetic  
field, 
$$\Bv(x) =
i 
\jj 
\int\ \frac{\vp(k)}{\sqrt{2\omega(k)}} (k \wedge  e_j(k)) \lk 
e^{-i\kx}\add(k,j)- 
e^{i\kx} a(k,j)\rk dk,$$
and we denote
$$\Bv=\Bv(0)$$
as an operator on $\fff$. 
Then, first working out the square in \kak{11}, one obtains 
\eq{14}
UHU\f =\frac{1}{2m}(p-\pf-e\Av)^2-\frac{\aa}{2m}\si\cdot \Bv+\hf.
\en
In \kak{14} $p$ is  multiplication by $k$ in the representation from 
\kak{12} and \kak{13}.
Thus the fibering is 
$$\hhh=\int_\BR^\oplus \hhh_p dp$$
with $\hhh_p$ isomorphic to $\BC^2\otimes\fff$ and 
$$UHU\f=
\int_\BR^\oplus \hp dp.$$
In the following we will regard $p\in\BR$ simply as a parameter. 
We also choose units such that $m=1$. 
Then the operator under study is 
\eq{hp}
\hp=\half\lk \p- \pf-e  \Av\rk^2-\frac{e}{2}\si \cdot \Bv+ \hf
\en
acting on 
$$\hhh_p=\BC^2\otimes \fff.$$
For simplicity the index $p$ of $\hhh_p$ will be omitted.

For $\aa=0$ \kak{hp} reduces to the noninteracting Hamiltonian 
\eq{16}
\hpz=\half(p-\pf)^2+\hf.
\en 
Clearly, with the definitions \kak{01} and \kak{02}, $\hpz$ is 
self-adjoint with the domain 
$D(\hf+\pf^2)=D(\hf)\cap D(\pf^2)$. The rest of \kak{hp} is regarded 
as the interaction part of the Hamiltonian, 
\eq{17}
\hi=\hp-\hpz 
=-{\aa}(\p-\pf)\cdot \Av+\frac{\at}{2}\Av^2-\frac{\aa}{2}\sigma\cdot 
\Bv.
\en 

For sufficiently small $\aa$, $\mm<e^\ast$,  $\hi$ is bounded 
relative to 
$\hpz$ with a bound less than 1, which by a theorem of Kato and 
Rellich implies that $\hp$ is a 
self-adjoint operator for every  $p\in\BR$. In addition $\hp$ is 
bounded from below. 
To see this let  $\ass(f)= \jj \int f(k,j) \ass(k,j)dk$. 
Using the inequalities 
$$\|a^\sharp (f) 
\psi\|\leq c_1 \lkk\jj \int \lk \omega(k)\f+1\rk  |f(k,j)|^2 dk\rkk^\hann 
\|(\hf+1)^\hann \psi\|,$$
$$\|a^\sharp (f) a^\sharp (f) \psi\|\leq  c_2 
\jj \int\lk \omega(k)\f+\omega(k)\rk|f(k,j)|^2 dk 
\|(\hf+1)\psi\|,$$
with some constants $c_1$ and $c_2$, 
one has 
\eq{co}
\|\hi\psi\|\leq \cco 
 \|(\hpz +1)\psi\|.
\en
Here 
$$\cco=\co \lkk \mm \lkk \qq \rkk^\hann\!\! \right.$$
$$
\left.\hspace{5cm} +\mm^2 \qqq\rkk$$
with some numerical constant $\co$ of order one. 
Thus  $\mm<\az$ with a suitable  $\az>0$ implies $\cco<1$. 

The goal of our paper is to study some ground state properties of 
$\hp$. 
The ground state energy of $\hp$ is 
$$E(p)=\inf\s(\hp)=\inf_{\psi\in D(\hp), \|\psi\|=1} (\psi, 
\hp\psi).$$
It is easily seen that resolvent 
$(\hp-z)\f$ with $z\not\in \s(\hp)$ is continuous in both 
$\p$ and $\aa$ in the operator norm. 
Thus $\epr$ is continuous in both $\p$ and $\aa$. 
If $E(p)$ is an eigenvalue, the corresponding spectral projection is 
denoted by $\pg $. 
$\tr \pg $ is the degeneracy of  the ground state. 
The bottom of the continuous spectrum is denoted by $E_{\rm c}(p)$. 
Under the assumptions (A.1)--(A.3)  one knows that 
$$E_{\rm c}(p)=\inf_{k\in\BR}\lkk E(p-k)+\omega(k)\rkk,$$
see \cite{fr2,sp3}. 
Thus it is natural to set 
$$\D(p)=E_{\rm c}(p)-E(p)=\inf_{k\in\BR}\lkk 
E(p-k)+\omega(k)-E(p)\rkk.$$
Let $\aa=0$. Then $\D(p)>0$ for $|p|<p_0$ with some $p_0$ by (A.1). 
Thus,   
by the continuity of $E(p)$ mentioned above, $\{(p,\aa)\in \BR\times 
\RR| \D(p)>0\}\not=\emptyset.$
As our  main result we state
\bt{main3}
Suppose  $|e|<e_0$  and $\D(p)>0$. 
Then $\tr \pg =2.$
\et
We briefly comment on our assumptions. 
$\D(p)>0$ means that we assume, rather than prove, a spectral gap. If instead  one would merely impose the implicit but rather natural condition that 
\eq{nat}
E(p)\geq E(0), 
\en
then,   following the arguments in \cite{fr2},  
one concludes in case of the dispersion relation \kak{10} that $\D(p)>0$
for 
$|p|<\sqrt{3}-1$ and  arbitrary  $\aa\in\RR$. For a general dispersion 
relation satisfying (A.1) to (A.3) a corresponding bound can be 
established. 
For  the spinless Pauli-Fierz Hamiltonian  \kak{nat} is 
proven through a suitable variant of a  diamagnetic inequality. 
We did not succeed to establish \kak{nat} for the Hamiltonian 
\kak{hp}. 
Note that on physical grounds one expects $\D(p)>0$ for $|p|< \pc$ 
and 
$\D(p)=0$ for $|p|>\pc$ with a simultaneous loss of the ground state. 
For the Hamiltonian \kak{hp} with $\omega$ given by \kak{10} one has 
$\pc <\infty$ at $e=0$ and expects $\pc<\infty$ to persist for all 
$e\not=0$ provided the spatial dimension $d\geq 3$.

The constant $\aa_0$ does {\it not} depend on $m_{\rm ph}$. Thus, 
together with 
\kak{nat}, 
the domain of validity of Theorem \ref{main3} is independent  of 
$m_{\rm ph}$ and it may seem  that one could take the limit $m_{\rm 
ph}\rightarrow 0$. 
Of course, thereby $\D(p)\rightarrow 0$. 
Unfortunately, this will not work, since the Pauli-Fierz Hamiltonian 
is infrared divergent and the number of photons increases without 
bound  as $m_{\rm ph}\rightarrow 0$ \cite{ch}. 
The physical ground states are no longer in  Fock space. 
The only exception is $\p=0$, and one might hope 
to apply our method directly to 
the case $\p=0$ and  
$m_{\rm ph}=0$. This problem will be taken up in Section 3.

Let us indicate the strategy for the proof of 
Theorem \ref{main3}. From a pull-through formula one estimates 
the overlap of a ground state with 
the subspace $P_0\hhh=\BC^2\otimes\{\BC \Omega\}$. 
If $\mm$ is sufficiently small, the overlap is large which implies 
$\tr \pg <3$. For a lower bound we will derive the algebraic relation 
$P_0\pg P_0=aP_0$, 
$a>0$, which implies $\tr \pg\geq 2$. 

 The Hamiltonian \kak{hp} is invariant under rotations  with axis 
$\widehat{p}=p/|p|$. To understand the implications let us 
define the field angular momentum relative to the origin by  
$$\jf=-i \jj\int \add(k,j) (k\wedge \nabla_k) a(k,j)dk$$
and the helicity by 
$$\ssf=i\int \hk \lkk \add(k,2)a(k,1)-\add(k,1)a(k,2) \rkk dk. $$
Let $\n\in\BR$ be an arbitrary  unit vector, 
$\t\in\RR$, and let 
 $R=R(\n,\t)$ be the rotation around $\n$ through the angle $\t$. 
We note  that 
$$\ejt \hf \emjt=\hf,$$
$$\ejt \pf \emjt=R \pf,$$
$$\ejt \Av \emjt=R\Av .$$
Define the total angular momentum through 
$$J=\jf+\ssf+\half\si.$$
Then it follows that 
$$\eejt \hp \eemjt =H_{R\f \p}.$$
In particular, the ground state energy is rotation  invariant,  
\eq{ep}
E(p)=E(R\p). 
\en 
If one sets  $\n=\vP$, then 
\eq{ang}
e^{i\t  \pj} \hp e^{-i\t \pj}=\hp, 
\en 
which expresses the rotation invariance relative to $\vP$. 

Since 
$\s(\vP\cdot  (\jf+\ssf))={\Bbb Z}$ and $
\s(\vP\cdot \si ) =\{-1,+1\}$, 
it follows that 
$$\s(\pj)=\hz,$$
where $\hz$ is the set of half integers 
$\{\pm\hann,\pm3/2,\pm5/2,...\}$. 
By virtue of \kak{ang},  $\hhh$ and $\hp$ are decomposable 
as 
$$\hhh=\bigoplus_{z\in \hz} \hhh(z),$$
$$\hp=\bigoplus_{z\in \hz}   \hp(z).$$
Therefore, if $\tr \pg=2$, the ground states $\griii\in\pg\hhh$ can 
be chosen such that 
$\gri\in\hhh(z)$ and $\grii\in\hhh(z')$ for some $z,z'\in\hz$. 
\bt{main2} 
Suppose  $|e|<e_0$  and $\D(p)>0$. 
Then $\hp$ has two orthogonal  ground states, $\griii$, 
with the property  $\griii\in\hhh(\pm\hann)$. 
\et
 
\section{Spectral properties} 
\subsection{Upper bound} 
Let us denote the number operator 
by 
$$\nf=\jj\int\add(k,j)a(k,j)dk.$$
In what follows 
$\gr$  denotes an {\it arbitrary}  normalized ground state of $\hp$. 
We note that  $a(k,j)\psi$, $\psi\in D(\nf^\hann)$, is well defined   
and 
$$(a(k,j)\psi)^{(n)}(\kj)=\sqrt{n+1}\psi^{(n+1)}(k,j,\kj).$$
Moreover it follows that 
$\|a(k,j)\psi\|\leq \|\nf^\hann \psi\|$ and 
$$(\psi, \nf \phi)=\jj\int(a(k,j)\psi, a(k,j)\phi) dk.$$
\bl{fp}
Suppose $\D(p)>0$. 
Then 
$$(\gr, \nf\gr)\leq 
2 \at\int 
\frac{(|k|^2/4)+6E(p)}{(E(\p-k)+\omega(k)-E(p))^2}\frac{|\vp(k)|^2}{\omega(k)}dk. 
$$ 
\el
\proof 
By the pull-through formula 
$$a(k,j)\hp \gr=\lk \hpk +\omega(k)\rk a(k,j)\gr$$
$$-\aa\sa \lkk\PP \cdot e_j(k)+\half \si\cdot (ik\wedge 
e_j(k))\rkk\gr . $$
Hence we have 
$$
a(k,j)\gr=\aa \sa \lk \hpk+\omega(k)-E(p) \rk\f $$
$$\times \lkk \PP \cdot e_j(k)+\half\si\cdot (ik\wedge 
e_j(k))\rkk\gr.$$
Thus 
$$(\gr, \nf \gr)=\jj \int \|a(k,j)\gr\|^2dk 
$$
$$\leq 
\jj \int\left|\sa\frac{\|\PP \cdot e_j(k)\gr+(\hann) \si\cdot 
(ik\wedge e_j(k))\gr\|}{E(\p-k)+\omega(k)-E(p)}\right|^2dk.$$
We note that 
$$\|\PP \cdot e_j(k)\gr\|^2=\|\mmm \si_\mu \PP_\mu  \si_\mu  
e_j^\mu(k)\gr\|^2$$
$$\leq 3 \mmm \|\si_\mu\PP_\mu \gr\|^2\leq 6  E(p), $$
and 
$$\|\si\cdot (ik\wedge e_j(k))\gr\|^2\leq 
|k|^2 \mmm \|(i\hk\wedge e_j(k))_\mu \gr\|^2\leq |k|^2. $$
Thus 
$$
\|\PP \cdot e_j(k)\gr+(\hann )\si\cdot (ik\wedge e_j(k))\gr\|^2\leq 
2\lkk (|k|^2/4)+6E(p)\rkk, $$
which leads  to 
$$(\gr, \nf\gr)\leq 2 \at\int 
\frac{(|k|^2/4)+6E(p)}{(E(\p-k)+\omega(k)-E(p))^2}\frac{|\vp(k)|^2}{\omega(k)}dk$$

and  the lemma follows.
\qed
We set 
$$\t(p)=\t(p,\aa)=2 \int 
\frac{(|k|^2/4)+6E(p)}{(E(\p-k)+\omega(k)-E(p))^2}\frac{|\vp(k)|^2}{\omega(k)}dk.$$

Note that $\t(p)$ is rotation  invariant, i.e. $\t(Rp)=\t(p)$ for an 
arbitrary  rotation $R$. 
Let $P_0=1\otimes P_\Omega$ be the projection onto $\BC^2\otimes 
\{\BC \Omega\}$. 

\bl{35}
Let 
$\D(p)>0$. 
Suppose 
$\mm <1/\sqrt{3 \t(p)} $.
Then $\tr \pg \leq 2$. 
\el
\proof
By Lemma \ref{fp} we  have $\tr(\P\nf)\leq \at \t(p)\tr \P.$
Therefore 
$$\tr \P-\tr(\P P_0)=\tr\P(I-P_0)\leq \tr(\P \nf)\leq \at \t(p) \tr 
\P,$$
and 
$$(1-\at \t(p))\tr \P\leq \tr (\P P_0)\leq 2,$$
which implies 
$$\tr\P\leq \frac{2}{1-\at \t(p)}<3.$$
Thus the lemma  follows.
\qed

\subsection{Lower bound}
We say that $\psi\in\fff$ is {\it real}, 
if for all $n\geq 0$,   $\psi^{(n)}$ 
is a real-valued function on 
$L^2(\RR^{3n}\times\{1,2\})$.  
The set of  real $\psi$ is denoted by $\fffr$.  
We define the set of reality-preserving operators $\rpof$ 
as follows:
$$\rpof=\left\{A| A:\fffr\cap D(A)\longrightarrow \fffr\right\}.$$
It is seen that 
$\hf$ and $\pf$ are in  $\rpof$. 
Since for all $k\in \RR$ and $z\in\RR$ with $z\not\in\s(\hpz)$,  
$$((\hpz-z)^k\psi)^{(n)}\nm$$
$$=
\lk \half\lk \p-\sumi k_i\rk^2+\sumi \omega(k_i) 
+z\rk^k
\psi^{(n)}\nm,$$
$(\hpz-z)^k$ is also in  $\rpof$. 
Moreover $\aaa(f)$ and $\add(f)$ are in  $\rpof$ for real $f$'s. 
In particular $\Av$ and $i\Bv$ are in $\rpof$.
Note that, 
if $\psi\in D(\Bv)\cap \fffr$, then 
$(\psi,\Bv \psi)=0.$
Let 
$\ffff=\bigcup_{N=0}^\infty \oplus_{n=0}^N \lk L^2(\BR\times\{1,2\})
\rk_{\rm sym}^n$ 
 denote the finite particle subspace of $\fff$. 
Note that $\Av$, $\Bv$,  $\zes$ and $\zess$ leave $\ffff$ invariant. 

\bl{real}
Suppose 
$\mm<\az$.
Let $x\in\ct$. 
Then there exists $a(t)\in\RR$ independent of $x$  such that for 
$t\geq 0$
\eq{eht}
(\xo, e^{-t
(\hp-\epr)
} \xo)_\hhh=a(t)(x,x)_{\ct}.
\en
\el
\proof 
Note that 
$\|\hi(1+\hpz)\f\|<1$ for $\mm<\az$ by \kak{co}. 
Then, by spectral theory, one has 
$$
\ehtt=\ln\left(1+\frac{t}{n} (\hp-\epr) \right)^{-n}$$
$$
=\lnk \left\{
\sum_{k=0}^m\ress \! \!
\left\{\left(-\frac{t}{n}\hib\right)\ress\right\}^k
\right\}^n 
$$
$$=\lnk 
\left\{
\res\left(
\sum_{k=0}^m  \lk-\frac{t}{n} \hiw\rk ^k     \right)
\res\right\}^n.$$
Here 
$$\hiw=\hiiw+i\si\cdot\BB,$$
$$\hiiw=\res(\hii-\epr) \res,$$
$$\BB=\res (i \Bv) \res,$$
$$\hii=-\aa(p-\pf)\cdot\Av+\frac{\at}{2}\Av^2.$$
It is seen that 
$$\hiw^2=
\hiiw \hiiw- \BB\cdot\BB +i\si\cdot (\hiiw 
\BB+\BB\hiiw-\BB\wedge\BB)=M+i\si\cdot L.$$
Here 
both of 
$M=\hiiw \hiiw- \BB\cdot\BB$ and 
$L=\hiiw \BB+\BB\hiiw-\BB\wedge \BB$ are  in  $\rpof$. 
Moreover 
$$\hiw^3=
\hiiw M-\BB L+i\si\cdot(\BB M+\hiiw L-\BB\wedge  L),$$
where 
both of $\hiiw M-\BB L$ and $\BB M+\hiiw L-\BB\wedge  L$ are 
also in   $\rpof$. 
Thus, repeating above procedure, one obtains 
$$\sum_{k=0}^m\lk-\frac{t}{n} \hiw\rk ^k=a_m+i\si\cdot b_m,$$
where $a_m$ and $b_m$ are in  $\rpof$.
Hence there exist   $a_{nm}\in \rpof$ and $b_{nm}\in\rpof$ such that  
$$\left\{\res\left(\sum_{k=0}^m   
\left(-\frac{t}{n} \hiw\rk ^k\right)\res\right\}^n=a_{nm}+
i\si\cdot b_{nm}.$$
Finally 
$$(\xo,\ehtt\xo)$$
$$=\lnk(x,x)(\Omega,a_{nm}\Omega)+
i\lnk(x,\si  x) (\Omega, b_{nm}\Omega).$$
Since the left-hand side is real, 
the second term of the right-hand side vanishes and 
$a(t)=\lnk(\Omega, a_{nm}\Omega)$ exists, 
which establishes  the desired result. 
\qed
\bl{po}
Suppose that $\D(p)>0$ and $\mm< 1/\sqrt{\t(p)}$. 
Then 
$(\gr,  P_0 \gr)\not =0.$ 
\el
\proof
Since $P_\Omega+\nf\geq 1$, we have from Lemma \ref{fp}  
$$(\gr,  P_0 \gr)\geq \|\gr\|^2-\| (1\otimes \nf^\hann)  \gr\|^2>1-\at 
\t(p)>0.$$
Thus the lemma follows. 
\qed
\bl{projection}
Suppose $\mm<\az$ and $\mm<1/\sqrt{\t(p)}$. 
Then there exists  $a>0$  such that 
\eq{contr}
P_0\pg   P_0=aP_0.
\en 
\el
\proof
Note that 
$$\pg  =s-\lim_{t\rightarrow\infty}\ehtt.$$
Thus by Lemma \ref{real}, 
$$(\xo,\pg  \xo)=\lim_{t\rightarrow\infty}(\xo,\ehtt\xo)
=\lim_{t\rightarrow\infty}a(t)(x,x)$$
for all $x\in\ct$. Since by Lemma \ref{po}, 
$(\xo,\pg  \xo)\not=0$ for some $x\in\BC^2$, 
$\lim_{t\rightarrow\infty}a(t)$ exists and it does not vanish. 
For arbitrary $\phi_1,\phi_2\in\hhh$, the polarization identity 
leads  to  
$(\phi_1,P_0\P P_0\phi_2)=a(\phi_1,P_0\phi_2).$ 
The lemma follows.
\qed
\bl{low}
Suppose $\mm<\az$ and $\mm<1/\sqrt{\t(p)}$. 
Then $\tr \pg  \geq 2$.
\el
\proof 
Suppose  $\tr \pg  =1$. 
Then $P_0\pg P_0/\tr P_0\pg P_0$ is a one-dimensional projection 
which contradicts \kak{contr}.
Thus the lemma follows.
\qed
\subsection{Proofs of Theorems  \ref{main3} and \ref{main2}} 
We define 
$$e_0= \inf \lkk \mm \,  \left| \mm<1/\sqrt{3\t(p)}, 
\mm<\aa^\ast\right.\rkk.$$
\noindent
{\it Proof of Theorem \ref{main3}:} \\ 
Since  $\D(p)>0$ and $\mm<e_0$, 
we conclude   $\tr \pg  \geq 2$ by Lemma \ref{low} and 
$\tr \pg  \leq 2$ by Lemma \ref{35}. 
Hence the theorem follows.
\qed
\noindent
{\it Proof of Theorem \ref{main2}:} \\ 
Without restriction in generality we may suppose $\vP=(0,0,1)$.  
Let $\friii$ be ground states of $\hp$ such that 
$\fri\in\hhh(z)$ and $\frii\in\hhh(z')$ with some $z,z'\in\hz$. 
$\ppo=\lk\begin{array}{c}\Omega\\ 0\end{array}\rk$ and 
$\mmo=\lk\begin{array}{c} 0 \\ \Omega\end{array}\rk$ are ground 
states of $\hpz$ and 
$\ppoo\in\hhh(\pm\hann)$. 
Let 
$\pg \ppo=c_1\fri+c_2\frii$
and 
$\pg \mmo=c_3\fri+c_4\frii$ with some $c_j\in\BC$ and 
 $Q_{\pm\hann}$ be the projection of $\hhh$ onto $\hhh(\pm\hann)$. 
Since  $P_0\pg P_0=a P_0$ we conclude that 
\eq{aa}
(\ppo, \pg \ppo)=a>0, 
\en
\eq{aaa} 
(\mmo,\pg\mmo)=a>0.
\en
Then $Q_\hann \pg \ppo\not=0$ and 
$Q_{-\hann} \pg\mmo \not=0$. 
The alternative 
$Q_\hann\fri\not=0$ or $Q_{\hann}\frii\not=0$ holds by \kak{aa}, 
the alternative  
$Q_{-\hann}\fri\not=0$ or $Q_{-\hann}\frii\not=0$ by \kak{aaa}. 
We may set  $Q_\hann \fri\not=0$. 
Then $\fri\in\hhh(\hann)$ and  $\frii\in\hhh(-\hann)$. 
\qed

\section{Zero total momentum}
\subsection{Spinless Hamiltonian}
In the spinless case $\hp$  simplifies to 
$$\hp=\half\PP^2+\hf$$ 
acting on $\fff$. The bound \kak{nat} is available and 
$\hp$ has at least one ground state for $|p|<p_{\rm c}$ with some $\pc>0$ 
and arbitrary $\aa$. 
To show the 
uniqueness of the ground state only the pull-through argument seems to be available. 
The details of Section 2.1 remain unchanged and one concludes that 
if 
$$\mm^2\leq \half \lkk 
\int\frac{E(p)}{(E(p-k)+\omega(k)-E(p))^2}\frac{|\vp(k)|^2}{\omega(k)}dk\rkk\f,$$
then $\tr \pg\leq 1$, which implies that 
the ground state of $\hp$ is unique.

\subsection{Confining potentials}
For $p=0$ no infrared divergence is expected and for the remainder of 
this section we set 
$$\omega(k)=|k|$$
as the physical dispersion relation. We did not 
succeed to apply the methods of Section 2 to this case. 
Therefore rather than considering $p=0$ directly we add to the 
Hamiltonian 
\kak{11} 
a confining potential, $V:\BR\rightarrow \RR$, 
which in spirit amounts to the same physical situation. The 
Hamiltonian under study is 
\eq{31}
\hv=\half\lkk\si\cdot\lk-i\nabla_x-e\Av(x)\rk \rkk^2+V(x)+\hf.
\en 
Let $$\pa=-\half\Delta+V$$ on $\LR\otimes\BC^2$, 
$$\sp=\inf\s_{\rm ess}(\pa), $$
$$\epp=\inf\s(\pa)$$ and 
$$E=\inf\s(\hv).$$ 
Let $V$ be relatively bounded with respect to  $-\Delta$ 
with a  bound  less than 1. 
Then $\hv$ is self-adjoint on $D(\Delta)\cap D(\hf)$ 
as  established in \cite{hi3}. 
For arbitrary $\aa$ the existence of ground states has been proven  
by 
Griesemer, Lieb, and Loss \cite{gll} under the condition that 
$\pa $ has  a ground state   separated by a gap from the continuous 
spectrum, i.e.
\eq{gap}
\gap>0.
\en 
We also refer to \cite{bfs3} for prior results, where in particular 
it is proven 
that 
 the charge density of an arbitrary ground state $\gr$ is localized, 
i.e.  
\eq{ex}
\|e^{c|x|}\gr\|\leq c_1 \|\gr\|
\en 
with some constant $c$.  
In the spinless case the uniqueness of the ground state, $\tr\pg=1$, 
would follow from a positivity argument \cite{hi9}. 
Let $\pel$ be the projection of the subspace spanned by ground states 
of $\pa$. 
Suppose \kak{gap}. Take $\aa$ such that 
\eq{33}
\sp-E>0,
\en 
which  can be satisfied  by the continuity of $E$ in  $\aa$. 
Then 
a pull-through argument and \kak{ex} yield that 
\eq{pl}
(\gr,(1\otimes\nf+\pel^\perp\otimes P_0)\gr)\leq 
\ee, 
\en 
where $\lime \ee=0$. 
Hence 
in the similar way as Lemma \ref{35} 
with $P_0$ and $\nf$ replaced by $\pel\otimes P_0$ and $1\otimes 
\nf+\pel^\perp\otimes P_0$, respectively,   
we see that  
if, in addition to \kak{33},  
$\aa$ satisfies 
$
\ee<1/3$, 
then 
\eq{2}
\tr \pg < 3.
\en 
The realness argument of Section 2.2 requires some extra conditions 
on $V$. 
\bt{main4}
Suppose  $V(x)=V(-x)$, $\sp-\epp>0$,  and $\tr\pel=2$. 
Then there exists a positive constant 
$e_{00}$ such that, 
if $\mm<e_{00}$, then  $\hv$ has a two-fold degenerate ground state. 
\et
\proof
Suppose that $\aa$ satisfies \kak{33} and $\ee<1$. 
\kak{pl} yields 
\eq{th}
(\gr, \pel\otimes P_0\gr)\geq 
\|\gr\|^2-(\gr,(1\otimes \nf+\pel^\perp\otimes P_0)\gr)
\geq 1-\ee>0.
\en 
Let $F$ denote 
the Fourier transform of $\LR$ and define the unitary operator of 
$\hhh$ 
by 
$T=F e^{ix\cdot\pf}$. Then we have 
$$T\hv T\f=\half\lkk \si\cdot (x-\pf-\aa \Av(0))\rkk^2+FVF\f +\hf.$$
The assumption $V(x)=V(-x)$ implies that $F V F\f$ is a reality 
preserving operator 
on $\LR$. 
Let $\varphi_{\rm el}$ be the ground state of $\pa$. 
In the similar way as Lemmas \ref{real} and \ref{projection} 
with $\Omega$ and $P_0$ replaced by 
$\varphi_{\rm el}\otimes\Omega$ and $\pel\otimes P_0$, respectively, 
it is established that by \kak{th} there exists a positive constant 
$c^\ast$ such that  if $\mm<c^\ast$, then \kak{33} and 
$\ee<1$ hold, 
 and 
$\tr\pg\geq 2$. 
Take $e_{00}=\sup\lkk \mm|  \ee<1/3,  \mm<c^\ast\rkk.$
Then the theorem follows from  \kak{2}.
\qed

\noindent
{\footnotesize 
{\bf Acknowledgment}

We thank Volker Bach for explaining to us that the overlap with 
the vacuum yields an upper bound on the degeneracy.
F.H. gratefully acknowledges the kind  hospitality at Technische
Universit\"{a}t M\"{u}nchen. This work is in part supported by the
Graduiertenkolleg  ``Mathematik in ihrer Wechselbeziehung zur Physik'' of
the LMU  M\"{u}nchen and Grant-in-Aid 13740106 for Encouragement of Young 
Scientists from the Ministry of Education, Science, Sports, and Culture.

}
\end{document}